\documentclass[mtplusscr,dvips]{arxstspdf}
\usepackage{dcolumn,mathrsfs,euscript}
\usepackage{graphicx}
\usepackage{flushend}
\usepackage{stfloats}
\usepackage{breakurl}

\volume{23}
\issue{3}
\pubyear{2008}
\firstpage{354}
\lastpage{364}
\doi{10.1214/08-STS260}

\makeatletter

\newcolumntype{d}[1]{D{.}{.}{#1}}
\DeclareMathAlphabet\mathcaligr{OMS}{cmsy}{m}{n}
\makeatother

\begin{document}
\begin{frontmatter}

\title{The Banff Challenge: Statistical Detection of a Noisy Signal}
\runtitle{Statistical Detection of a Noisy Signal}

\begin{aug}
\author[a]{\fnms{A. C.} \snm{Davison}\corref{}\ead[label=e1]{Anthony.Davison@epfl.ch}} and
\author[b]{\fnms{N.} \snm{Sartori}\ead[label=e2]{sartori@unive.it}}
\runauthor{A. C. Davison and N. Sartori}

\affiliation{Ecole Polytechnique F\'ed\'erale
de Lausanne, Universit\`a ``Ca' Foscari''
Venezia}

\address[a]{Anthony Davison is Professor of
Statistics, Institute of Mathematics,
Ecole Polytechnique F\'ed\'erale
de Lausanne (EPFL), Lausanne, Switzerland (\printead{e1}).}
\address[b]{Nicola Sartori is Assistant Professor of
Statistics, Dipartimento di Statistica, Universit\`a ``Ca' Foscari''
Venezia, Venezia, Italy and SSAV (\printead{e2}).}

\end{aug}

%
\begin{abstract}
Particle physics experiments such as those run in the Large Hadron
Collider result in huge quantities of data, which are boiled down to a
few numbers from which it is hoped that a signal will be detected.
We discuss a simple probability model for this and derive frequentist
and noninformative Bayesian
procedures for inference about the signal. Both are highly
accurate in realistic cases, with the frequentist procedure having the
edge for interval estimation,
and the Bayesian procedure yielding slightly better point estimates. We also
argue that the significance, or $p$-value, function based on the
modified likelihood root provides a comprehensive presentation of the
information in the data and should be used for inference.

\end{abstract}

%
\pdfkeywords{Bayesian inference,
higher-order asymptotics,
Large Hadron Collider,
likelihood,
noninformative prior,
orthogonal parameter,
particle physics,
Poisson distribution,
signal detection}

\end{frontmatter}

\section{Introduction}

Particle physics experiments such as those conducted in the Large
Hadron Collider entail the detection of a signal in the presence of
background noise.
This essentially statistical topic has been discussed intensively in
the recent literature
(\cite{Mandelkern2002}, \cite{FraserReidWong2004}, and the references
therein)
and at a series of meetings
involving statisticians and physicists; see \citet{Lyons2008} for
more details and
further references. One key issue is the setting of confidence limits
on the
underlying signal, based on data from independent observation channels.

In the simplest version of the problem there is just one channel, the
observation from which is the number of times a particular event in a
particle accelerator has been observed. This is supposed to have a
Poisson distribution with mean $\gamma\psi+\beta$, where the positive
known constants
$\beta$ and $\gamma$ represent respectively
a background rate at which the event occurs and the efficiency of the
measurement device. There is a\vadjust{\goodbreak} substantial physical literature about
inference for the focus of interest, the unknown parameter $\psi$.
Typically frequentist inference is preferred to Bayesian approaches,
but this is the subject of a lively debate among the scientists
involved. In order to compare properties of various procedures for
inference about $\psi$, it was decided at the workshop on \textit
{Statistical Inference Problems in High Energy Physics and Astronomy}
held at the Banff International Research Station in 2006
that one participant would create artificial data that should mimic those
that might arise when the Large Hadron Collider is running, and that
other participants would
attempt to set confidence limits for the known underlying signal.
Thus was the Banff Challenge
(\url{http://newton.hep.upenn.edu/\textasciitilde heinrich/birs/}) born.

For a single channel the challenge may be stated as follows: the
available data $y_1,y_2,y_3$ are assumed to be realizations of
independent Poisson random variables with means
$\gamma\psi+\beta, \beta t, \gamma u$, where $t,u$ are known
and the parameters $\psi,\beta,\gamma$ are unknown. This expands the
formulation above
to allow for uncertainty about the values of the background $\beta$
and the efficiency $\gamma$, which are supposed to be estimable from
subsidiary experiments of known lengths $t$ and $u$.
The goal is to summarize the evidence concerning
$\psi$, large estimates of which will suggest presence of the signal.
The parameters $\beta$ and $\gamma$ are necessary for realism, but
their values are only of
concern to the extent that they impinge on\vadjust{\goodbreak} inference for
$\psi$.

This is a highly idealized version of one of many statistical problems
that will arise
in dealing with data from the Large Hadron Collider. The model is very
simple, but important inferential issues arise nonetheless: how is
evidence about the
value of $\psi$ best summarized? How should one deal with the nuisance
parameters
$\beta,\gamma$? This second issue is even more critical in the case of
multiple channels, where the number of
nuisance parameters is much larger.
Below we follow \citet{FraserReidWong2004} in arguing that
the evidence concerning $\psi$ is best summarized through a so-called
significance function, and
in Section \ref{liktheory} describe the general construction of
significance functions that yield highly accurate frequentist
inferences even with many nuisance parameters; such a significance
function is equivalent to a set
of confidence intervals at various levels. In Section \ref{likelihood} we
give results for the Poisson model for the two cases laid out in the
Banff Challenge, with one channel and with ten channels.

Statisticians are in broad agreement that the likelihood function
is central to parametric inference. Bayesian inference uses the
likelihood to update prior information to give a posterior probability
density that summarises what it is reasonable to believe
about the parameters in light of the data
(\cite{Jeffreys1961}, \cite{ForsterOHagan2004}). This approach is
attractive and widely used in applications, but
scientists with different priors may arrive at different conclusions
based on the same data.
One might argue that this is inevitable given the varied points of view
held within any scientific community,
but this lack of uniqueness is awkward when an objective statement is
sought. One way to unite this multiplicity of
possible posterior beliefs is to base inference on a noninformative
prior, which we discuss
in Section \ref{bayes} for the Poisson model described
above.

One aspect we discuss only peripherally is the choice of the Poisson
distribution to represent the variation of the observed events.
Statisticians typically regard a model as one of many possibilities,
whereas physicists tend to argue from first principles and the known
properties of the systems that they study toward a strong belief that
certain models, such as the Poisson law used here, are correct. Under
the Banff Challenge the Poissonness of the observations is taken as given.\vspace*{-1pt}

\section{Likelihood and Significance} \label{liktheory}\vspace*{-1pt}

There are many published accounts of modern likelihood theory.
The outline below is based on \citet{BrazzaleDavisonReid2007},
wherein further references may be\vadjust{\goodbreak} found.

We consider a probability density function $f(y;\psi,\break\lambda)$ that
depends on two parameters. The interest parameter $\psi$ is the focus
of the investigation; one may wish to test whether it has a specific
value $\psi_0$, or
to produce a confidence interval for the true but unknown value of
$\psi
$. Often $\psi$ is scalar,
as here: $\psi$ represents the signal central to our enquiry. The
nuisance parameter
$\lambda$ is not of direct interest, but must be included for the model
to be realistic. In the single-channel case the vector $\lambda=(\beta
,\gamma)$ represents the background signal and
measurement efficiency. Let $\theta=(\psi,\lambda)$ denote the entire
parameter vector.

The log likelihood function is defined as $\ell(\theta) = \log
f(y;\theta)$.
The maximum likelihood estimator $\widehat\theta$ satisfies $\ell
(\widehat\theta
)\geq\ell(\theta)$ for all $\theta$ lying in the parameter space~$\Omega_\theta$,
which we take to be an open subset of $\mathbb{R}^{d}$.
We suppose that $\psi$ may take values in the interval $(\psi_-,\psi
_+)$, where one or both of the limits
$\psi_-,\psi_+$ may be infinite. A natural summary of the support for
$\psi$ provided by the
combination of model and data is the profile log likelihood
\[
\ell_{\mathrm{p}}(\psi) = \ell(\widehat\theta_\psi) = \ell(\psi
,\widehat\lambda_\psi
) = \max_\lambda\ell(\psi,\lambda),
\]
where $\widehat\lambda_\psi$ is the value of $\lambda$ that
maximizes the
log likelihood for fixed $\psi$.

Under regularity conditions on $f$ under which a~random sample
of size $n$ is generated from $f(y;\theta_0)$,
the estimator $\widehat\theta$ has an approximate normal
distribution with
mean $\theta_0$ and variance matrix $j(\widehat\theta)^{-1}$, where
$j(\theta)= -\partial^2\ell(\theta)/\partial\theta\,\partial\theta
^\mathrm{T}$ is
the observed information matrix.
This result can be used as the basis of confidence intervals for $\psi
_0$, based on the limiting
standard normal, $\mathcal{N}(0,1)$, distribution of the Wald pivot
$t(\psi_0) = j_{\mathrm{p}}(\widehat\psi)^{1/2}(\widehat\psi-\psi
_0)$, where
\[
j_{\mathrm{p}}(\psi) = -{\partial^2\ell_{\mathrm{p}}(\psi)\over\partial\psi
^2} =
{|j(\psi,\widehat\lambda_\psi)|\over|j_{\lambda\lambda}(\psi
,\widehat\lambda
_\psi)|} ,
\]
$|\cdot|$ indicates determinant, and $j_{\lambda\lambda}(\theta)$
denotes the $(\lambda,\lambda)$ corner of the observed information
matrix. In many ways a preferable
basis for confidence intervals is the likelihood root
\[
r(\psi) = \operatorname{sign}(\widehat\psi-\psi)[2\{
\ell_{\mathrm{p}}(\widehat\theta
)-\ell_{\mathrm{p}}(\widehat\theta_\psi)\}]^{1/2},
\]
which may also be treated as an $\mathcal{N}(0,1)$ variable. If it is required
to test the hypothesis that $\psi=\psi_0$ against the
one-sided hypothesis that $\psi>\psi_0$, then the quantities $1-\Phi
\{
r(\psi_0)\}$ and
$1-\Phi\{t(\psi_0)\}$ are treated as significance probabilities, also
known as $p$-values,
small values of which will cast doubt on the belief that $\psi=\psi_0$.
Throughout the paper $\Phi$
represents the cumulative probability function of the standard normal
distribution.

The monotonic decreasing function $\Phi\{r(\psi)\}$ is an example of a
significance function,
from which we may draw
inferences about $\psi$. An approximate lower confidence bound $\psi
_\alpha$ for $\psi_0$ is the solution to the equation $\Phi\{r(\psi
)\}
= 1-\alpha$; the confidence interval $(\psi_\alpha,\psi_+)$ should
contain $\psi_0$ with probability $1-\alpha$. An approximate upper
bound $\psi_{1-\alpha}$
is obtained by solution of $\Phi\{r(\psi)\}=\alpha$, giving confidence
interval $(\psi_{-},\psi_{1-\alpha})$, and the two-sided interval
$(\psi
_\alpha,\psi_{1-\alpha})$ will contain $\psi_0$ with probability
approximately
$(1-2\alpha)$. Using these so-called first-order approximations, these
one-sided intervals in fact contain $\psi_0$
with probability $1-\alpha+ \mathcal{O}(n^{-1/2})$, while the two-sided
interval contains $\psi_0$ with probability
$(1-2\alpha)+\mathcal{O}(n^{-1})$. Significance functions may be
based on the
Wald pivot $t(\psi)$ or on related quantities involving the log
likelihood derivative $\partial\ell/\partial\psi$, which also have
approximate $\mathcal{N}(0,1)$ distributions for large $n$, but the intervals
based on $r(\psi)$ are preferable because they always yield confidence
sets that are subsets of $(\psi_-,\psi_+)$. Further, they are invariant
to invertible interest-preserving reparametrization, of the form
$(\psi,\lambda) \mapsto(g(\psi),h(\lambda,\psi))$: if $\mathcaligr{I}$ is
a confidence
interval for $\psi$ in the original parametrization, then $g(\mathcaligr{I})$ is
the corresponding interval in the new parametrization; this property is
not possessed by
intervals based on the Wald pivot, for example.

A large body of literature on higher-order parametric asymptotics, both
Bayesian and frequentist, has converged on a few key formulae that are
useful for inference. There are
numerous derivations of these formulae in different cases, for example
by Laplace approximation to posterior densities or by saddlepoint
approximation to conditional densities; see \citet{Reid2003} or
\citeauthor{Davison2003}
(\citeyear{Davison2003}, Sections 11.3.1, 12.3.3). Fuller accounts are given by
\citet
{BrazzaleDavisonReid2007}, \citet{Severini2000}, \citet
{PaceSalvan1997} and \citet{BarndorffNielsenCox1994}.
Perhaps the
most practicable route to these improved inferences is through
significance functions based on the modified likelihood root
%
%
\begin{equation}
\label{rstar}
r^*(\psi) = r(\psi) +{1\over r(\psi)} \log\biggl\{{q(\psi)\over
r(\psi
)}\biggr\},
\end{equation}
where
%
%
\begin{equation}
\label{inference.Q.eq}\qquad
q(\psi)= {| \varphi(\widehat\theta) - \varphi(\widehat
\theta_\psi) \varphi_\lambda(\widehat\theta_\psi)|\over
| \varphi_\theta(\widehat\theta)|}
\biggl\{ {|j(\widehat\theta)|\over|j_{\lambda
\lambda}(\widehat
\theta_\psi)|}\biggr\}^{1/2}
\end{equation}
is determined by a local exponential family approximation whose
canonical parameter $\varphi(\theta)$ is described below, and
$\varphi
_\theta$ denotes the $d\times d$ matrix $\partial\varphi/\partial
\theta
^\mathrm{T}$ of partial derivatives.
The numerator of the first term of (\ref{inference.Q.eq})
is the determinant of a $d\times d$ matrix whose first column
is $\varphi(\widehat\theta) - \varphi(\widehat\theta_\psi)$
and whose remaining
columns are $\varphi_\lambda(\widehat\theta_\psi)$. For continuous
variables, one-sided
confidence intervals based on the significance function $\Phi\{
(r^*(\psi
)\}$ have coverage error $\mathcal{O}(n^{-3/2})$ rather than $\mathcal
{O}(n^{-1/2})$.

For a sample of independent continuous observations $y_1,\ldots, y_n$,
we define
\[
\varphi(\theta)^\mathrm{T}= \sum_{k=1}^n {\partial\ell
(\theta;y)\over
\partial y_k}\bigg|_{y=y^0} V_k,
\]
where $y^0$ denotes the observed data, and $V_1,\ldots, V_n$ is a set
of $1\times d$ vectors that depend
on the observed data alone. If the observations are discrete, then the
theoretical accuracy of the approximations is\break reduced to $\mathcal{O}(n^{-1})$,
and the interpretation of\break significance functions such as $\Phi\{
r^*(\theta)\}$ changes\break slightly. In the discrete setting of this paper
we take
(\citeauthor{DavisonFraserReid2006},~\citeyear{DavisonFraserReid2006})
%
%
\begin{equation}
\label{inference.V.eq2}
V_k = {\partial\mathrm{E}(Y_k;\theta)\over\partial\theta
^\mathrm{T}}
\bigg|_{\theta=\widehat\theta},
\end{equation}
where $\mathrm{E}$ denotes expectation.
An important
special case is that of a log likelihood with independent contributions
of curved exponential
family form,
\index{exponential family!curved}%
%
%
\begin{equation}
\label{inference.L.eq}
\ell(\theta) = \sum_{k=1}^n \{\alpha_k(\theta) y_k -
c_k(\theta
)\},
\end{equation}
where $\alpha_k(\theta) y_k$ denotes scalar product. In this case
%
%
\begin{equation}\label{inference.V.eq2a}
\varphi(\theta)^\mathrm{T}= \sum_{k=1}^n\alpha_k(\theta) V_k.
\end{equation}

Inference using (\ref{rstar}) is easily performed. If functions are
available to compute
$\ell(\theta)$ and $\varphi(\theta)$, then the maximizations needed
to obtain
$\widehat\theta$ and $\widehat\theta_\psi$ and the
differentiation needed to
compute (\ref{inference.Q.eq})
may be performed numerically.

Inferences based on (\ref{rstar}) are invariant to addition to the log
likelihood of quantities dependent only on the data, which lead to
affine transformations of $\varphi(\theta)$ by quantities that are
parameter independent and which therefore leave (\ref{inference.Q.eq})
unchanged.

As with other uses of approximations in applied mathematics, asymptotic
results like those sketched above in which $n\to\infty$ are intended
for use with samples whose size is fixed and finite. The key is that
some measure of information, which may depend on the parameter values
as well as on sample size, becomes large; in the present case
information also accumulates as the Poisson means increase. Both
general theory and the simulations described below suggest that the
higher-order
approximations outlined above
are highly accurate even when little information is available.

\section{Likelihood Inference}
\label{likelihood}

\subsection{Model Formulation}

Under the proposed model, the observation for the $k$th channel is
assumed to be a realization of $Y_k=(Y_{1k},Y_{2k},Y_{3k})$, where the
three components are independent Poisson variables with respective
means $(\gamma_{k} \psi+\beta_{k},\beta_{k} t_k,\gamma_{k} u_k)$,
for $k=1,\ldots,n$.
Here $Y_{1k}$ represents the main measurement, $Y_{2k}$ and $Y_{3k}$
are respectively subsidiary background and efficiency measurements, and
$t_k$ and $u_k$ are known positive constants.

The signal parameter $\psi$ is of interest, and $(\beta_{1},\gamma
_{1},\break\ldots,\beta_{n},\gamma_{n})$ is treated as a nuisance parameter.
In principle the nuisance parameters are positive and $\psi\geq0$, but
it is mathematically reasonable to entertain negative values for $\psi
$, provided $\psi>\max_k\{-\beta_{k}/\break \gamma_{k}\}$. Below we use this
extended parameter space for numerical purposes, but restrict
interpretation of the results to the physically meaningful values
$\psi\geq0$, as suggested by \citet{FraserReidWong2004}.

For computational purposes we take $\lambda=(\lambda_{11},\lambda
_{21},\break\ldots,\lambda_{1n},\lambda_{2n})$, with
$(\lambda_{1k},\lambda_{2k})=(\log\beta_{k}-\log\gamma_{k},\break\log
\beta_{k})$,
so that $\exp(\lambda_{1k})> -\psi$ and $\lambda_{2k}\in\mathbb{R}$,
$k=1,\ldots,n$. The invariance properties
outlined in the previous section imply that inferences on $\psi$ are
unaffected by this reparametrization.

The log likelihood function for $\theta=(\psi,\lambda)$ has curved
exponential family form
(\ref{inference.L.eq}) with
%
%
\begin{eqnarray} \label{alphai}
\alpha_k(\theta)^\mathrm{T}
&=&\{\log(\psi e^{\lambda
_{2k}-\lambda
_{1k}} +e^{\lambda_{2k}} ),\nonumber\\
&&\hspace*{32.6pt}\lambda_{2k}, (\lambda
_{2k}-\lambda_{1k})\} , \nonumber\\[-8pt]\\[-8pt]
y_k^\mathrm{T}&=& (y_{1k},y_{2k},y_{3k}) ,\nonumber \\
c_k(\theta) &=& (\psi+u_k) e^{\lambda_{2k}-\lambda_{1k}} +
(1+t_k)
e^{\lambda_{2k}} . \nonumber
\end{eqnarray}
In general,\vspace*{2pt} $\widehat\theta$ and $\widehat\theta_\psi$ must be computed
numerically. It is convenient to compute $\widehat\theta_\psi$
first, and
then obtain $\widehat\theta$ by maximizing the
profile log likelihood $\ell(\widehat\theta_\psi)$.

The dimension of the nuisance parameter may be reduced by a
conditioning argument
that applies to Poisson responses, but for simplicity of exposition we
use the Poisson
formulation here. The trinomial model that emerges from the
conditioning is used below in
Section~\ref{bayes.model}. Properties of the Poisson model imply that
numerical results from the two
formulations are identical.

\subsection{One Channel} \label{onechannel}

When data from only one channel are available, that is, $n=1$, the log
likelihood has full exponential form. The canonical parameter $\varphi
(\theta)$ given by (\ref{alphai}) is then equivalent to (\ref
{inference.V.eq2a}) in the sense that any affine transformation of the
canonical parameter gives the same $q(\psi)$ in (\ref{inference.Q.eq})
and the same inference for $\psi$.

A standard way to summarize the evidence concerning $\psi$ is to
present the profile log likelihood $\ell_{\mathrm{p}}(\psi)$ and the
significance function $\Phi\{r(\psi)\}$ (\cite{FraserReidWong2004}),
but, as mentioned above, more accurate inferences are obtained from the
modified likelihood root, $r^*(\psi)$. As the profile log likelihood
equals $-r(\psi)^2/2$, the quantity $-r^*(\psi)^2/2$ can be
regarded as the adjusted profile log likelihood corresponding to the
significance function
$\Phi\{r^*(\psi)\}$.

\begin{figure*}

\includegraphics{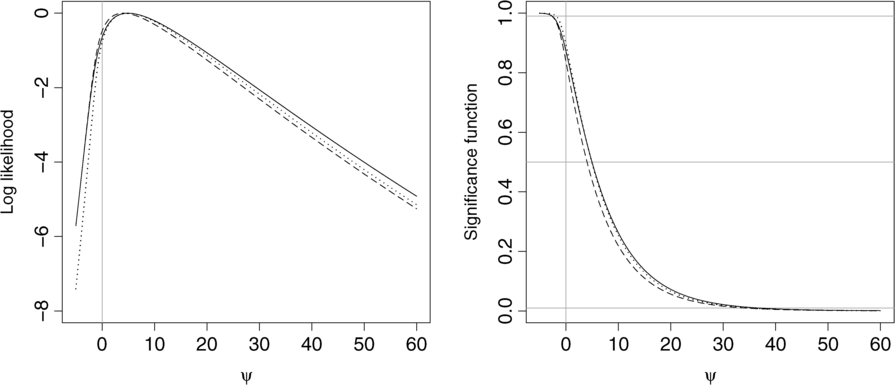}

\caption{Inferential summaries for the illustrative single-channel data.
Left panel: profile relative log likelihood $\ell_{\mathrm{p}}(\psi)-
\ell_{\mathrm{p}}(\widehat\psi)$ (dashes), $-r^*(\psi)^2/2$ (solid) and
$-r_B^*(\psi)^2/2$ (dots). Right panel: $\Phi\{r(\psi)\}$ (dashes),
$\Phi\{r^*(\psi)\}$ (solid) and $\Phi\{r^*_B(\psi)\}$ (dots).
Horizontal lines are at values $0.99$, $0.01$ and $0.5$, and give
respectively the lower and upper bounds of a confidence
interval of level 0.98, and a median unbiased estimate of $\psi$. The
intersection of a significance function with the vertical line at $\psi
=0$ gives the corresponding $p$-value for
testing the hypothesis $\psi=0$ against $\psi>0$.}
\label{fig1}
\end{figure*}

For illustration we consider data with $y_1=1$, $y_2=8$, $y_3=14$ and
$t=27,u=80$, for which
Figure \ref{fig1} shows the profile and the adjusted profile log
likelihoods and the corresponding significance functions, and a
Bayesian solution whose construction is explained in Section \ref{bayes}.
The maximum likelihood estimate, $\widehat\psi=4.021$, may be determined
from the significance function as the solution to the equation $\Phi\{
r(\widehat\psi)\}=0.5$. The analogous estimate obtained using the modified
likelihood root, the median unbiased estimate
$\widehat\psi^*=4.966$, satisfies $\Phi\{r^*(\widehat\psi^*)\}
=0.5$. The
corresponding estimator has equal probabilities of falling to the left
or to the right of the true parameter value, a property preferable to
classical unbiasedness because it does not depend on the parametrization.

%

One minus the value of the significance function at $\psi=0$ gives the
significance probability for testing the presence of a signal, namely
the $p$-value
for testing the hypothesis $\psi=0$ against the
one-sided hypothesis $\psi>0$. In the present example, $\Phi\{r(0)\}
=0.837$ and $\Phi\{r^*(0)\}=0.873$, thus giving $p$-values respectively
equal to $0.163$ and $0.127$, both weak evidence of a positive signal.
This is hardly surprising, as $y_1=1$: just one event has been observed.

As explained in Section \ref{liktheory}, the significance function provides
lower and upper bounds
for any desired confidence level. Figure \ref{fig1} indicates the
choice of lower and upper bounds for
level $0.99$. In particular, for the modified likelihood root, we get
$\Phi\{r^*(\psi^*_{0.01})\}=0.99$ and
$\Phi\{r^*(\psi^*_{0.99})\}=0.01$, with $\psi^*_{0.99}=-2.603$ and
$\psi^*_{0.01}=36.519$. It is possible for these limits to be negative,
as happens in the present case for the lower bound. In such instances,
we take as a limit the maximum $\max(\psi^*_\alpha,0)$ of the actual
limit, $\psi^*_\alpha$, and the lower physically admissible value of
zero. The fact that the lower bound is zero in this case is coherent
with the $p$-value for testing a positive signal. In fact, a right-tail
confidence interval of level $0.99$ in this case contains all possible
parameter values, also including $0$; thus it is $[0,+\infty)$. A
left-tail confidence interval is $[0,36.510)$, although its usual
interpretation makes it ill-suited to claim the presence of signal.
The analogous limits obtained using the likelihood root $r(\psi)$
are $\psi_{0.99}=-2.644$ and $\psi_{0.01}=33.835$.

In extreme situations confidence limits at any standard choice of
$\alpha$ may be negative, thus giving confidence intervals including
only the value $\psi=0$. We see this feature of the method as a
perfectly sensible frequentist answer (see also \cite{Cox2006},
Example~3.7). In such instances the $p$-value for testing $\psi=0$
against the alternative $\psi>0$ would be very close to $1$, thus
strongly suggesting that there is no positive signal. However, doubt is
cast on the model when no physically realistic
parameter value is supported by the observed data.

%
\begin{figure*}

\includegraphics{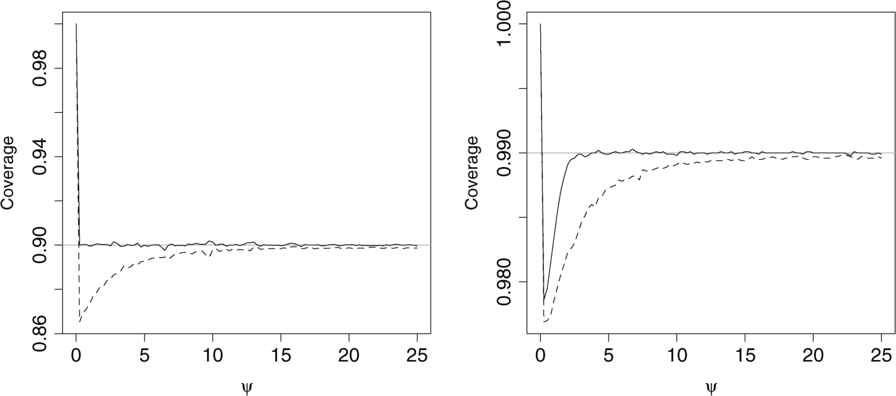}

\caption{Coverages of $0.90$ (left panel) and $0.99$ (right panel)
upper bounds from 39,700
simulated datasets from a single channel, with large uncertainty in the
nuisance parameters, from the Banff Challenge. The solid and dashed
lines correspond respectively to $r^*(\psi)$ and $r_B^*(\psi)$.
The ideal coverage is shown by the horizontal lines.}
\label{fig2}
\end{figure*}

In the Banff Challenge only coverage of left-tail confidence intervals
(upper bounds) was tested,
though we regard $p$-values and lower bounds as more appropriate for
inference on $\psi$. Figure \ref{fig2} shows the coverage of $0.90$
and $0.99$ confidence limits as functions of $\psi$ for a set of 39,700
simulated datasets with large variability in the values of the nuisance
parameters. The coverage is very good, with only minor undercoverage in
the $0.99$ upper bounds when the parameter $\psi$ is small. Similar
results were obtained for another set of simulated datasets, with
smaller variability in the nuisance parameters. We also performed some
simulation studies with a variety of parameter
values, and found that our procedure is typically highly accurate.
Table \ref{tab0} displays results in the worst scenario that we found.
Apart from
some minor issues in the right tail, $r^*$ performs extremely well.

In some boundary cases with $y_1=0$ it is impossible to compute the quantities
needed for (\ref{inference.Q.eq}). In these rare cases we replaced
$r^*(\psi)$ with $r(\psi)$.

\subsection{Several Channels}\label{multichannel}

Our approach extends easily to multiple channels. When there are $n>1$
channels, the nuisance parameters $(\lambda_{1k},\lambda_{2k})$ are
channel-specific, so the profile log likelihood is simply the sum of
profile log likelihood contributions for the individual channels, which
is then maximized numerically to get the overall estimate $\widehat
\theta
=(\widehat\psi,\widehat\lambda)$.

The remaining ingredient needed to compute the modified likelihood root
$r^*(\psi)$
is the $2n+1$-dimen\-sional canonical parameter $\varphi(\theta)$, which
can be obtained using~(\ref{inference.V.eq2a}) and (\ref
{inference.V.eq2}). The first element of $\varphi(\theta)$ is
\[
\sum_{k=1}^n e^{\widehat\lambda_{2k}-\widehat
\lambda_{1k}}\log
(\psi e^{\lambda_{2k}-\lambda_{1k}}+e^{\lambda_{2k}}) ,
\]
and the $2n$ other elements are
\begin{eqnarray}
&&\widehat\psi e^{\widehat\lambda_{2k}-\widehat\lambda_{1k}}
\log(\psi e^{\lambda_{2k}-\lambda_{1k}}+e^{\lambda
_{2k}})\nonumber\\
&&\quad{}
+u_j (\lambda_{2k}-\lambda_{1k}) e^{\widehat\lambda
_{2k}-\widehat
\lambda_{1k}} ,\nonumber\\
&&e^{\widehat\lambda_{2k}}
\log(\psi e^{\lambda_{2k}-\lambda_{1k}}+e^{\lambda
_{2k}})
+t_j \lambda_{2k} e^{\widehat\lambda_{2k}} ,\nonumber\\
&&\eqntext{k=1,\ldots, n.}
\end{eqnarray}
Any affine transformation of $\varphi(\theta)$ would give the same
modified likelihood root.

%
\begin{table}[b]
\caption{Empirical coverage probabilities in a single-channel
simulation with 10,000 replications,
$\psi=1$, $\log\beta=1.1$, $\log\gamma=0$, $t=33$ and $u=100$}
\label{tab0}
\begin{tabular*}{\tablewidth}{@{\extracolsep{\fill}}cccc@{}}
\hline
\textbf{Probability} & $\bolds r$ & $\bolds{r^*}$ & $\bolds{r^*_B}$ \\
\hline
0.0100 & \textbf{0.0080} & 0.0092 & 0.0104 \\
0.0250 & 0.0225 & 0.0253 & 0.0263 \\
0.0500 & \textbf{0.0437} & 0.0500 & 0.0514 \\
0.1000 & \textbf{0.0887} & 0.0995 & 0.1019 \\
0.5000 & \textbf{0.4669} & 0.5054 & 0.5045 \\
0.9000 & 0.8947 & 0.9051 & 0.9036 \\
0.9500 & \textbf{0.9186} & 0.9461 & \textbf{0.9320} \\
0.9750 & 0.9736 & \textbf{0.9809} & \textbf{0.9785} \\
0.9900 & \textbf{0.9816} & \textbf{0.9816} & \textbf{0.9816} \\
\hline
\end{tabular*}
\tabnotetext[]{}{Figures
in bold differ from the nominal level by more than
simulation error.}
\end{table}

%
\begin{figure*}

\includegraphics{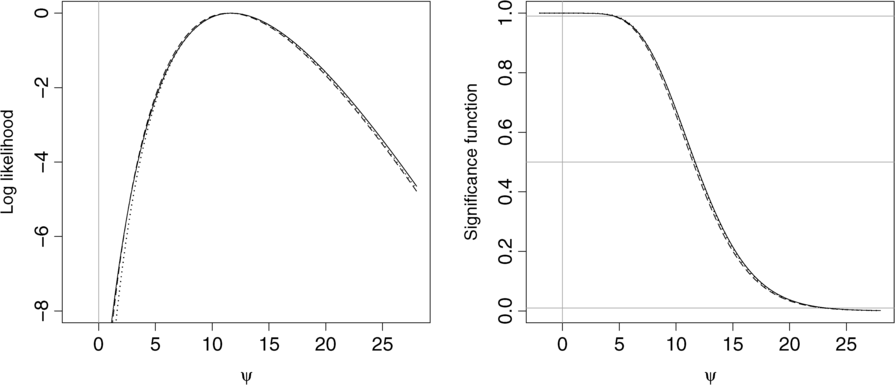}

\caption{Inferential summaries for the simulated multiple-channel data
in Table \protect\ref{tab1}.
For details, see caption to Figure \protect\ref{fig1}.}
\label{fig3}
\end{figure*}

Figure \ref{fig3} gives the profile and adjusted profile log
likelihoods for $\psi$ and the corresponding significance functions for
an illustrative dataset with $n=10$ channels shown in Table \ref{tab1}.
The interpretation of these plots is the same as for Figure \ref{fig1}.
The modified likelihood root gives a $p$-value of $7.709\times10^{-7}$
for testing the presence of a signal, whereas that based on the
likelihood root is $3.124\times10^{-7}$.
The estimates are $\widehat\psi^*=11.682$ and $\widehat\psi
=11.487$ and the
lower and upper bounds are $\psi^*_{0.99}=4.572$,
$\psi^*_{0.01}=23.191$ and $\psi_{0.99}=4.496$,
$\psi^*_{0.01}=22.907$. There is strong evidence of a positive signal
from these data, though the modified likelihood root $r^*(\psi)$ gives
weaker support than does the ordinary likelihood root
$r(\psi)$. In fact the evidence here corresponds to significance near
to the ``$5\sigma$'' level used by particle physicists when deciding
whether or not to announce a discovery (\cite{Lyons2008}).

%
\begin{table}[b]
\caption{Simulated multiple-channel data}
\label{tab1}
\begin{tabular*}{\tablewidth}{@{\extracolsep{\fill}}d{2.0}cd{2.0}d{2.0}cc@{}}
\hline
\multicolumn{1}{@{}c}{\textbf{Channel}} & \multicolumn{1}{c}{$\bolds{y_1}$}
& \multicolumn{1}{c}{$\bolds{y_2}$} & \multicolumn{1}{c}{$\bolds{y_3}$}
& \multicolumn{1}{c}{$\bolds{t}$} & \multicolumn{1}{c@{}}{$\bolds{u}$} \\
\hline
1 & 1 & 7 & 5 & 15 & 50 \\
2 & 1 & 5 & 12 & 17 & 55 \\
3 & 2 & 4 & 2 & 19 & 60 \\
4 & 2 & 7 & 9 & 21 & 65 \\
5 & 1 & 9 & 6 & 23 & 70 \\
6 & 1 & 3 & 5 & 25 & 75 \\
7 & 2 & 10 & 10 & 27 & 80 \\
8 & 3 & 6 & 12 & 29 & 85 \\
9 & 2 & 9 & 7 & 31 & 90 \\
10 & 1 & 13 & 13 & 33 & 95 \\
\hline
\end{tabular*}
\end{table}

Boundary samples also arise in the multiple-channel case, though less
frequently than with a single channel. In such cases we again used the
likelihood root $r(\psi)$ for inference on $\psi$.

Figure \ref{fig4} shows coverages of the $0.90$ and $0.99$ left-tail
confidence intervals (upper bounds) computed with the modified
likelihood root from 70,000 simulated datasets with $n=10$ from the
Banff Challenge. Our approach seems to perform satisfactorily even with
as many as 20 nuisance parameters,
though there is again some undercoverage for small values of $\psi$.
Table \ref{tab2} reports coverage probabilities for limits at various
confidence levels for a simulation
performed with $\psi=2$. The results for the modified likelihood root
are always within simulation error of the nominal levels, thus giving
very accurate inference for~$\psi$.

\section{Bayesian Inference}
\label{bayes}

\subsection{Noninformative Priors}

There is a close link between the modified likelihood root and
analytical approximations useful for Bayesian inference.
Suppose that posterior inference is required for $\psi$ and that the
chosen prior density is $\pi(\psi,\lambda)$. Then it turns out that
replacing (\ref{inference.Q.eq}) with
\[
q_B(\psi) = \ell'_{\mathrm{p}}(\psi) j_{\mathrm{p}}(\widehat\psi)^{-1/2}
\biggl\{ {|j_{\lambda\lambda}(\widehat\theta_\psi)
|\over
|j_{\lambda\lambda}(\widehat\theta)|}\biggr\}^{1/2}
{\pi(\widehat\theta)\over\pi(\widehat\theta_\psi)}
\]
in formula (\ref{rstar}), where $\ell'_{\mathrm{p}}$ is the derivative of
$\ell_{\mathrm{p}}(\psi)$ with respect to $\psi$,
leads to a Laplace-type approximation to the marginal posterior
distribution for $\psi$, that we will denote by $r^*_B(\psi)$. This may
be used to include prior information, but, as mentioned above, the
choice of prior density can be vexing.
In this section we discuss noninformative Bayesian inference for $\psi$.

%
\begin{figure*}

\includegraphics{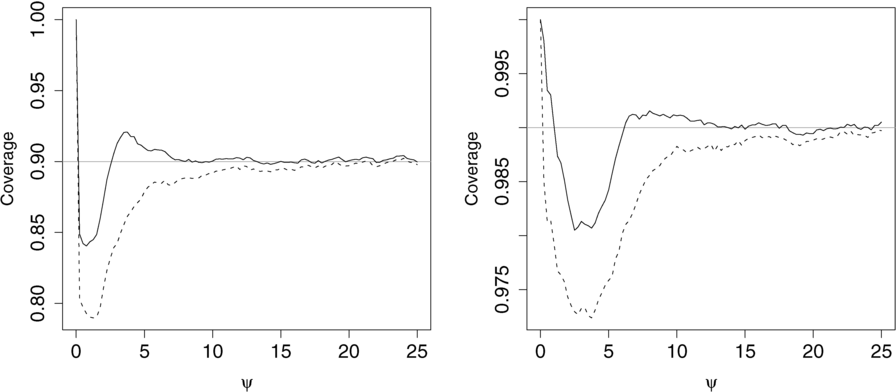}

\caption{Empirical coverages of $0.90$ (left panel) and $0.99$ (right
panel) upper bounds from 70,000 simulated multiple-channel datasets
from the Banff Challenge. The solid and dashed lines correspond
respectively to $r^*(\psi)$ and $r_B^*(\psi)$.}\vspace*{-3pt}
\label{fig4}
\end{figure*}

For models with scalar $\psi$ and a nuisance parameter $\xi$ that is
orthogonal to $\psi$ in the sense of \citet{CoxReid1987},
\citet
{Tibshirani1989} shows that up to a certain degree of approximation, a
prior density that is noninformative about $\psi$ is proportional to
%
%
\begin{equation}
\label{tibs}
|i_{\psi\psi}(\psi,\xi)|^{1/2} g(\xi)  \,d\psi \,d\xi,
\end{equation}
where $i_{\psi\psi}(\psi,\xi)$ denotes the $(\psi,\psi)$ element
of the
Fisher information matrix, and
$g(\xi)$ is an arbitrary positive function that satisfies mild regularity
conditions. Under further mild
conditions (\ref{tibs}) is a Jeffreys prior for $\psi$, and it is also
a matching
prior: following \citet{WelchPeers1963}, \citet
{ReidMukerjeeFraser2002}
show how (\ref{tibs}) yields
$(1-\alpha)$ one-sided Bayesian posterior confidence intervals that
contain $\psi$ with
probability $(1-\alpha)+\mathcal{O}(n^{-1})$ in a frequentist sense.
Unfortunately (\ref{tibs})
requires one to express the model in terms of an orthogonal
parametrization, and this may be
impossible. Below we rewrite it in terms of an arbitrary parametrization.

Suppose therefore that the model is parametrized in terms of a scalar
interest parameter
$\psi$ and a column vector nuisance parameter $\zeta=\zeta(\psi,\xi)$,
with the log likelihood written as $\ell^*\{\psi,\zeta(\psi,\xi)\}
=\ell
(\psi,\xi)$. Then the elements of the Fisher information matrices in
the two parametrizations are related by the equations
%
%
\begin{eqnarray}
\label{info}
i_{\psi\psi} &=& i^*_{\psi\psi}+2\zeta_\psi^\mathrm{T}i^*_{\zeta
\psi} + \zeta
_\psi^\mathrm{T}i^*_{\zeta\zeta}\zeta_\psi,\nonumber\\
i_{\xi\psi} &=& \zeta_\xi^\mathrm{T}i^*_{\zeta\psi} + \zeta_\xi
^\mathrm{T}i^*_{\zeta
\zeta} \zeta_\psi,\\
i_{\xi\xi} &=& \zeta_\xi^\mathrm{T}i^*_{\zeta\zeta} \zeta_\xi,\nonumber
\end{eqnarray}
where $i_{\xi\psi}=\mathrm{E}(-\partial^2\ell/\partial\xi
\partial\psi^\mathrm{T})$,
$i^*_{\zeta\zeta} = \mathrm{E}(-\partial^2\ell^*/\break\partial\zeta
\partial\zeta^\mathrm{T})$,
$\zeta_\psi= \partial\zeta/\partial\psi$, and so forth, with
%
\begin{table}[b]
\caption{Empirical coverage probabilities in a multiple-channel
simulation with 10,000 replications,
$\psi=2$, $\beta=(0.20, 0.30, 0.40, \ldots, 1.10)$, $\gamma=(0.20,
0.25, 0.30, \ldots, 0.65)$, $t=(15, 17, 19, \ldots,33)$ and $u=(50, 55,
60, \ldots, 95)$}
\label{tab2}
\begin{tabular*}{\tablewidth}{@{\extracolsep{\fill}}cccc@{}}
\hline
\textbf{Probability} & $\bolds{r}$ & $\bolds{r^*}$ & $\bolds{r^*_B}$ \\
\hline
0.0100 & 0.0099 & 0.0101 & 0.0109 \\
0.0250 & 0.0244 & 0.0255 & 0.0273 \\
0.0500 & 0.0493 & 0.0519 & 0.0542 \\
0.1000 & 0.0967 & 0.1012 & 0.1035 \\
0.5000 & \textbf{0.4869} & 0.5043 & 0.5027 \\
0.9000 & \textbf{0.8900} & 0.9013 & 0.8942 \\
0.9500 & \textbf{0.9421} & 0.9499 & \textbf{0.9427} \\
0.9750 & \textbf{0.9687} & 0.9759 & \textbf{0.9689} \\
0.9900 & \textbf{0.9875} & 0.9913 & \textbf{0.9864} \\
\hline
\end{tabular*}
\tabnotetext[]{}{Figures in bold differ from the nominal level by more than
simulation error.}
\end{table}
$\mathrm{E}$
again denoting expectation. Parameter orthogonality implies
that $i_{\xi\psi}\equiv0$, so provided $\zeta_\xi$ is not
identically zero,
$\xi=\xi(\psi,\zeta)$ is determined by the partial differential equation
%
%
\begin{equation}
\label{orthog}
\zeta_\psi= - i^{*-1}_{\zeta\zeta}i^*_{\zeta\psi},
\end{equation}
which always has a set of solutions for scalar $\psi$.
On substituting (\ref{orthog}) into the first expression in (\ref
{info}), we find that in terms of the original parametrization the
required element of the Fisher information matrix may be written as
\[
i_{\psi\psi} = i^*_{\psi\psi} - i^*_{\psi\zeta}i^{*-1}_{\zeta
\zeta
}i^{*}_{\zeta\psi},
\]
whence the noninformative prior (\ref{tibs}) may be written as
%
%
\begin{eqnarray}
\label{prior}
&&| i^*_{\psi\psi} - i^*_{\psi\zeta}i^{*-1}_{\zeta\zeta
}i^{*}_{\zeta\psi}|^{1/2}\nonumber\\[-8pt]\\[-8pt]
&&\quad{}\cdot
g\{\xi(\psi,\zeta)\} |\partial\xi/\partial\zeta|
  d\psi d\zeta,\nonumber
\end{eqnarray}
which requires that the orthogonal parameter
$\xi$ be expressed in terms of the original parameters; cf. expression
(5) of \citet{Tibshirani1989}.
In the next section we derive (\ref{prior}) for the single- and
multiple-channel models of Section \ref{likelihood}.

\subsection{Application to Poisson Model}
\label{bayes.model}

The single-channel model may be reparametrized in terms of $\psi$,
$\gamma$ and $\zeta=\beta/\gamma$, in which case $Y_1,Y_2,Y_3$ are
independent Poisson variables with means $\gamma(\psi+\zeta),\zeta
\gamma t,\gamma u$. This implies that the trinomial density of
$(Y_1,Y_2,Y_3)$ conditional on the total $S=Y_1+Y_2+Y_3$ does not
depend on
$\gamma$, and there is no
loss of information on $\psi$ and $\zeta$ if we base inference on the
trinomial or more generally the multinomial model
(\cite{BarndorffNielsen1978}, Chapter 10). In particular, frequentist
inferences on $\psi$ based on the original model or on the conditional
trinomial model lead to exactly the same results.
Here $\zeta$ is scalar.
Apart from additive constants, the corresponding log likelihood is
\begin{eqnarray*}
\ell^*(\psi,\zeta) &=& y_1\log(\psi+\zeta) + y_2\log\zeta\\
&&{} - s\log
(\psi
+\zeta+u + \zeta t),\quad
\psi+\zeta,\zeta>0,
\end{eqnarray*}
and $\mathrm{E}(Y_1\mid S=s) = s(\psi+\zeta)/\pi$, $\mathrm
{E}(Y_2\mid S=s) = s t \zeta
/\pi$, where
$\pi=\psi+\zeta+u+ \zeta t$. Thus in this parametrization
the Fisher information matrix for the trinomial model has form
\begin{eqnarray*}
&&i^*(\psi,\zeta)\\
&&\quad = {s\over\pi^2(\zeta+\psi)}\\
&&\qquad{}\cdot \pmatrix{u+\zeta t &
u-\psi
t \cr u-\psi t&
\{\psi t (\psi+u) + \zeta u (1+t)\}/\zeta},
\end{eqnarray*}
and the orthogonal parameter is a solution of the equation
\[
\zeta_\psi= \zeta(\psi t- u)/\{\psi t (\psi+u) + \zeta u (1+t)\},
\]
such as
\begin{eqnarray*}
\xi(\psi,\zeta) &=& t\log\zeta+ \log(\zeta+\psi) \\
&&{}- (1+t)\log
(\psi+\zeta
+u + \zeta t).
\end{eqnarray*}
It is impossible to express $\zeta$ explicitly as a function of $\psi$
and $\xi$, and hence to use the noninformative prior in the form
(\ref
{tibs}), but (\ref{prior}) is readily obtained,
and after a little algebra turns out to be proportional to
%
%
\begin{eqnarray}
\label{single.prior}\quad
&& \biggl[{\psi t (\psi+u) + \zeta u (1+t)\over\zeta^2(\zeta+\psi)^2
(\psi
+\zeta+u + \zeta t)^3}\biggr]^{1/2}\nonumber\\
&& \quad{}\cdot
g\biggl\{{(\zeta+\psi)\zeta^t\over(\psi+\zeta+u + \zeta
t)^{1+t}}
\biggr\}
\,d\psi \,d\zeta,\\
&&\eqntext{\zeta,\psi+\zeta>0,}
\end{eqnarray}
for an arbitrary but smooth and positive function $g$.

If data $(y_{1k},y_{2k},y_{3k},t_k,u_k)$ are available for $n$
independent channels, then the conditioning
argument above
yields $n$ independent trinomial distributions for
$(y_{1k},y_{2k},y_{3k})$ conditional on the
$s_k=y_{1k}+y_{2k}+y_{3k}$, whose probabilities depend on the
parameters $\psi, \zeta_k$.
Apart from an additive constant the log likelihood is
\begin{eqnarray*}
&& \ell^*(\psi,\zeta_1,\ldots, \zeta_n)\\
&&\quad = \sum_{k=1}^n
\{y_{1k}\log(\psi+\zeta_k)\\
&&\hspace*{38.5pt}{} + y_{2k}\log\zeta_k -
s_k\log(\psi+\zeta_k+u_k + \zeta_k t_k)\},
\end{eqnarray*}
where $\psi>-\min(\zeta_1,\ldots,\zeta_n)$ and $\zeta_1,\ldots
,\zeta_n>0$.
Calculations like those leading to (\ref{single.prior}) reveal that
the noninformative prior for $\psi$ is proportional to
%
%
\begin{eqnarray}
\label{several}
&&\Biggl|\sum_{k=1}^n {s_kt_ku_k}/(\zeta_k+\psi+u_k+\zeta t_k)\nonumber\\
&&\quad\hspace*{5.6pt}{}\cdot
\{\psi(\psi+u_k)t_k +
\zeta_ku_k(1+t_k)\}\Biggr|^{1/2}\\
&&\quad{}\cdot
\prod_{k=1}^n {\psi(\psi+u_k)t_k + \zeta_k u_k(1+t_k)
\over\zeta_k(\zeta_k+\psi)(\zeta_k+\psi+u_k+\zeta_k t_k)},\nonumber
\end{eqnarray}
times an arbitrary function of the quantities
\begin{eqnarray}
\xi_k(\psi,\zeta_k) &=& t_k\log\zeta_k + \log(\zeta_k+\psi)\nonumber\\
&&{} -
(1+t_k)\log(\psi+\zeta_k+u_k + \zeta_k t_k),\nonumber\\
&&\eqntext{k=1,\ldots, n.}
\end{eqnarray}
Although (\ref{several}) depends on the data through $s_1,\ldots,s_n$,
these are constants under
the trinomial model, as are the $t_k$ and $u_k$ under both Poisson and
trinomial models.
The presence of $s_kt_ku_k$ in the first term of (\ref{several}) has
the heuristic explanation that
a channel for which this product is large will contain more information
about its nuisance
parameters.

\subsection{Numerical Results}

We first consider the single-channel data analyzed in Section \ref
{onechannel}, with $y_1=1$, $y_2=8$,
$y_3=14$, and $t=27$, $u=80$. The dotted lines in Figure \ref{fig1}
show the approximate posterior function, $-r^*_B(\psi)^2/2$, and the
corresponding significance function obtained using the noninformative
prior (\ref{single.prior}), with $g$ taken to be a constant function.

Typically the prior density yields larger lower\break bounds and smaller
upper bounds than those obtained from the frequentist solution, because
the effect of the prior is to inject information about the parameter
of interest. In the present case, the estimate
$\widehat\psi_B^*=4.9182$, which satisfies $\Phi\{r^*_B(\widehat
\psi_B^*)\}
=0.5$, is smaller than the corresponding estimate obtained using
$r^*(\psi)$, and the $0.99$ lower and upper bounds are respectively
given by $\Phi\{r^*_B(\psi^*_{B;0.01})\}=0.99$ and
$\Phi\{r^*_B(\psi^*_{B;0.99})\}=0.01$, with $\psi^*_{B;0.99}=-1.820$
and $\psi^*_{B;0.01}=35.094$.

The $p$-value for testing the hypothesis $\psi=0$\break against the
one-sided hypothesis $\psi>0$ is equal to $1-\Phi\{r^*_B(0)\}=0.1063$,
which is again a weak evidence of a positive signal.

The coverage properties of the noninformative\break Bayesian solution are
similar to but not
quite so good as those of the frequentist solution, as shown in Figure
\ref{fig2} and by the simulation results
reported in the last column of Table \ref{tab0}.

Similar behavior is seen in the multichannel case. Figure \ref{fig3}
shows the approximate posterior function, $-r^*_B(\psi)^2/2$, and the
corresponding significance function obtained using the noninformative
prior (\ref{several}) times a constant function of $\xi_k(\psi,\zeta
_k)$, $k=1,\ldots,n$,
for the data in Table \ref{tab1}.
The approximate Bayesian solution gives a $p$-value of $4.865\times
10^{-8}$ for testing the presence of a signal, smaller than that
obtained from the frequentist solutions in Section \ref{multichannel}.
The estimate is $\widehat\psi_B^*=11.632$ and the lower and upper bounds
are $\psi^*_{B;0.99}=4.699$ and $\psi^*_{B;0.01}=23.030$.
There is stronger evidence of a positive signal from this approach than
from the modified likelihood root $r^*(\psi)$ and the ordinary
likelihood root $r(\psi)$. However, simulation results reported in
Figure \ref{fig4} and Table \ref{tab2} show that the coverage of
confidence sets based on the approximate Bayesian solution is not
quite so good as for sets based on the modified likelihood root.

\section{Discussion}

In this paper we propose procedures based on modern likelihood theory
for detecting a signal
in the presence of background noise, using a simple statistical model.
We suggest the use of the significance function based on the modified
likelihood root as a comprehensive summary of the information for the
parameter given the model and the observed data, from which $p$-values
and one- or two-sided confidence limits can be obtained directly.

Even when there are 20 nuisance parameters, our
frequentist procedure appears to give essentially exact inferences for
the signal parameter $\psi$.
Its noninformative Bayesian counterpart performs slightly worse in
terms of coverage of
confidence intervals and levels for tests, but provides slightly better
point estimates as solutions to the equation $\Phi\{r^*_B(\psi)\}=0.5$,
analogous to median unbiased estimates. The most serious
departures from the correct coverage are for small values of $\psi$,
corresponding to weak signals,
and arise because in such cases very low counts $y_1$ corresponding to
the observed signal
are quite likely to arise. The case of a weak signal seems to be of
little practical interest,
because in such cases no strong significance can be obtained. Although
the Banff Challenge
concerned significance at the 90\% and 99\% levels, both general theory
and the accuracy of our
results suggest that similar precision can be expected for much more
extreme significance levels.

If $y_1=0$ our higher-order approaches break down, though a closely
related first-order inference is available. Such cases are
scientifically uninteresting, but to avoid difficulties it is tempting
to replace $y_1$ by $y_1+c$,
where $c$ is a small positive quantity. \citet{Firth1993}
investigates under what circumstances this modification yields
an improved estimate of the interest parameter in exponential family models,
taken on the canonical scale of the exponential family. Our model is
not a linear exponential family,
but ideas of \citet{Kosmidis2007} might be used to choose $c$ to yield
an improved estimate of $\psi$.
Our main interest is in confidence intervals and tests, however, and
since Firth's correction
corresponds to use of a default Jeffreys prior and we have found that
use of
a noninformative prior does not improve coverage properties of our
method, one should not be optimistic about the effect of this
correction in our context.

In some instances the method may lead to empty confidence intervals or
intervals including only the value $\psi=0$. Though galling to the
experimenters, this is not a critical problem from a frequentist
perspective. On the one hand,
even in such extreme samples the confidence function would yield a
$p$-value to test for the presence of a signal, and on the other hand,
the concentration of the likelihood and significance functions in a
region of physically meaningless values of the parameter might suggest
that the model is inappropriate.

\section*{Acknowledgments}
The work was supported by the Swiss
National Science
Foundation, the Italian Ministry of Education (PRIN 2006) and the
EPFL. We thank the organizers
of the Banff workshop for inviting us to take part, the participants
for stimulating discussions, and David
Cox, Rex Galbraith, two referees and the editor for comments on this paper.
We thank particularly Joel Heinrich for the computations underlying
Figures \ref{fig2} and \ref{fig4}.

\end{document}